\documentclass[numberedappendix]{emulateapj}
\usepackage{rotating}
\usepackage{amsmath,amssymb,amsbsy}
\usepackage{natbib}
\usepackage{graphicx}

\begin{document}
\title{The Minimum Description Length Principle and  
Model Selection in Spectropolarimetry}

\author{A. Asensio Ramos}
\affil{Instituto de Astrof\'{\i}sica de Canarias, 38205, La Laguna, Tenerife, Spain}
\email{aasensio@iac.es}

\begin{abstract}
It is shown that the two-part Minimum Description Length Principle can be used to
discriminate among different models that can explain a given observed
dataset. The description length is chosen to be the sum of the lengths of the 
message needed to encode the model plus the message needed to encode
the data when the model is applied to the dataset.
It is verified that the proposed principle can efficiently distinguish the model
that correctly fits the observations while avoiding over-fitting.
The capabilities of this criterion are shown in two simple problems
for the analysis of observed spectropolarimetric signals. The first is the de-noising 
of observations with the aid of the PCA technique. The second is the selection of 
the optimal number of parameters in LTE inversions.
We propose this criterion as a quantitative approach for distinguising the most
plausible model among a set of proposed models. This quantity is very easy
to implement as an additional output on the existing inversion codes.
\end{abstract}


\section{Introduction}
When a scientist tries to analyze a given observed data set, it is customary to 
begin by examining the data in various ways, such
as plotting the data and looking for patterns. In a second step, the scientist 
proposes a number of physically plausible models
that can reproduce the observed data set. A model fitting procedure is then applied 
to the data set so that the parameters
that characterize each model are inferred. After all the proposed models are fitted 
to the data, the next step is to compare them
and infer which of the fitted models is the most suitable. Methods for such a task 
have been developed. For instance, for models that represent a hierarchical
structure, we can use Akaike's Information Criterion \citep{akaike74} 
or Mallows' $C_p$ \citep{mallows73}. These methods can 
be applied even in the case that the set of proposed models does not
contain the perfect model. In this case, the aim is to select the most optimal
one. If the models are of completely different type and do not belong to a
hierarchical structure of models, cross-validation type methods can be applied.
However, they can be computationally demanding.

The standard procedure to select the optimum model is to make use of the 
Occam's Razor Principle (also known as the Principle of Parsimony). This
principle is commonly applied in science and is usually thought to be an heuristic 
approach for eliminating unnecessary complex hypothesis. The
principle states that the selected model has to provide an equilibrium between the 
model complexity and its fidelity to the
data. Nevertheless, determining the simplest model is often very complicated. 
It is usually argued that the number of parameters that parameterizes the  
model should be smaller than the number of degrees of freedom of the data. 
However, it is a difficult matter to estimate the number of degrees of freedom
of the data.

An alternative and successful procedure is to consider the problem in terms of a 
communication process. Assume that a sender S is interested in sending a given
observation to a receiver R. Several techniques are available for such a task. The 
most trivial one is to send the whole amount of data through a given channel. If
the data set is very large, the length of the message will be consequently very 
large. If S is able to model the observation using a given model, it is wiser and
shorter to transmit the model followed by the points in the observation that are not 
correctly reproduced by the model. If the model is good and simple enough, the
length of the message between S and R will be much shorter than sending
the complete observed data set. This compression will be degraded when the
model needed for explaining the observations is made unnecessarily complex.
\cite{rissanen78} was the first in suggesting that the code length could be
used for model comparison. This principle is nowadays known as the Minimum
Description Length (MDL) Principle, which states that we should \emph{choose 
the model that gives the shortest description of the data}.
In this framework, there is an interplay between models that are concise and  
easy to describe and models that produce a good
fit to the data set and captures the important features evident in the data.
Of course, this is neither the only nor the best strategy for such purposes. 
However, its application is desirable in comparison with ad-hoc or trial-and-error 
ways of performing model selection. The reason is that MDL principle 
has strong theoretical roots that lie on the Kolmogorov complexity theory
\citep{vitanyi00,gao00}. As we will show, the practical application of the
MDL principle is very easy to implement and it constitutes an ideal
approach for model selection.

This paper will be focused on one of the version of the MDL principle, the 
so-called two-part MDL. This strategy was developed by
Rissanen in a series of papers published during the seventies and eighties 
\citep[e.g.,][]{rissanen78,rissanen83,rissanen86}
and summarized in \cite{rissanen89}. We give a brief
summary of the main results in section \ref{sec:theory}. As shown below, the 
main idea is to write the description length of a given model applied
to a data set as the sum of the length of the code for describing the model and the 
length of the code for describing the data set
fitted by the model. 

The interpretation of spectropolarimetric data in solar and stellar observations 
allows to infer information about the properties of the magnetic field. Almost 
always, the recovery of the magnetic field vector is based on the assumption
of a model. These models depend on a given amount of variables that can be
estimated by fitting the observed Stokes profiles. Among these models, we can
find the Milne-Eddington approximation, the LTE approximation, the MISMA
hypothesis \citep{sanchez_almeida_misma96}, etc. Sometimes, it is clear 
that a model is the most appropriate when observational clues are available. 
However, it is customary that these clues are not present and one relies
on one model based on completely or partially subjective reasons. 

Some of these models are based on an extraordinarily large number of 
parameters. Until now, there is not any critical investigation toward 
analyzing whether enough information is available in the observations to
constraint such a large amount of parameters. This work is a first step
in this direction. We propose the application of the MDL criterion to 
quantitatively differentiate between possible models taking into account
the information available in the observations.

\section{Minimum Description Length}
\label{sec:theory}
\cite{rissanen78} related the problem of finite dimensional parameter estimation to 
the problem of designing an optimal encoding scheme.
We consider the problem of transmitting a set of data from a sender S to a 
receiver R. The sender must first communicate the
type of model that can be used for describing the data. Consider, for instance, a 
set of points that we want to fit with a polynomial model.
In this case, neither the order of the polynomial $k$, nor the value of the 
parameters of the polynomial $\{a_i,i=0,\ldots,k\}$ are known, so that this 
constitutes
an extremely ill-posed problem. However, the MDL principle can be used to find
the ``most plausible'' model that explains the observed
points. In the first step, the sender has to communicate the order of the polynomial 
$k$. When the sender and the receiver agree
on the type of model to use, the sender has to communicate the model itself, 
sending through the channel the $k$ parameters $a_k$. If noise is
present or if the proposed model is incomplete, the sender has to 
additionally communicate the deviations of the model from the data.

The question of simplicity can consequently be tackled by taking advantage 
of the work of \cite{rissanen78}. Each model is reduced to bits, the most
fundamental information unit. This allowed to transform the Occam's 
Razor principle into a completely functional
principle. The measure of simplicity is just the number of bits required to correctly
and univoquely transmit a set of observations by using a model. The sender S 
takes a set of observations as input, encodes and sends a message that contains 
all the information about the model and the data to the receiver R. Finally, the 
receiver decodes the message and produces an output. In 
information-preserving
encoding schemes, the output obtained by R has to be identical to the 
original observation performed by S. 
Let $D$ be a set of observations (dataset) and $M$ a model that is used to
describe them. The quantity $L(M)$ represents the length of the
code in bits necessary to encode the model $M$. As well, $L(D|M)$ represents
the length of the data encoded using the model $M$ (this term can be alternatively
seen as the residual between the data $D$ and the model $M$). The total length
of the message is:
\begin{equation}
\label{eq:length}
L = L(M) + L(D|M).
\end{equation}
The MDL principle tries to minimize $L$ and the model associated with this 
minimum length is selected as the most plausible model.

The previous analysis can be alternatively viewed from the Bayesian perspective
\citep[see, e.g.,][]{gao00}. 
The aim is to infer a model $M$ from a set of
observations $D$. The solution lies in choosing the model that maximizes the 
posterior 
probability $p(M|D)$, which can be expressed as
follows with the aid of the Bayes' theorem:
\begin{equation}
\label{eq:bayes}
p(M|D) = \frac{p(D|M) p(M)}{p(D)}.
\end{equation}
The term $p(D)$ can be considered as a normalizing factor and represents the 
probability that the data set $D$ occurs. The term $p(M)$
is the a priori probability of the model, that is, the probability that the model $M$ is 
true before any data set has been observed.
Finally, $p(D|M)$ is the likelihood of the data given the model $M$. The 
relationship between the MDL formalism given by Eq.
(\ref{eq:length}) and the Bayes formalism given by Eq. (\ref{eq:bayes}) is obtained
by making use of Shannon's optimal coding theorem
\citep{shannon48a,shannon48b}. This theorem states that the length of the ideal 
code for a value $x$ of a variable $X$ which follows from a known
probability distribution $p(X)$ is given by:
\begin{equation}
\label{eq:shannon}
L(x) = -\log p(X=x).
\end{equation}
Taking the negative logarithm of Eq. (\ref{eq:bayes}), we obtain:
\begin{equation}
\label{eq:bayes_log}
-\log p(M|D) = -\log p(D|M) - \log p(M) + \log p(D).
\end{equation}
The most plausible model is the one that minimizes $-\log p(M|D)$, that is, the one 
that minimizes $-\log p(D|M) - \log p(M)$.
Note that we have ignored the influence of $p(D)$ since it is a constant that is 
shared for all the models and it is only
associated with the data set. According to Shannon's theorem, 
the minimization of Eq. (\ref{eq:bayes_log}) is
equivalent to the minimization of Eq. (\ref{eq:length}).

\subsection{Code Length Formulae}
Shannon's theorem states the length of the optimal code needed for transmitting a 
given number $x$ whose probability distribution
is known. Although extremely important, the theorem can be useless because
this probability distribution is usually not known and only 
approximate values of the length of the message can be
obtained. An example is when one needs to transmit a set of integer or real 
numbers for which no probability distribution is known.
It has been demonstrated that knowing exactly the encoding scheme is
of accesory importance \citep{rissanen78}. What is fundamental to know is which
model gives the minimal length of the message given an arbitrary encoding 
scheme. In the following, we assume that the probability distribution is not
known and we present existing estimations for the length of the message needed 
for communicating 
integer and real numbers. These formulae represent
the first estimations that were obtained for the lengths of the message for 
communicating models \citep{rissanen78}. Since they are not
based on any probability distribution for the integer or real numbers, they are 
known as universal priors \citep{rissanen78}.

Assume that we want to encode an integer number $n$ in its binary 
representation. The binary representation of $n$ has
a length that can be estimated to be of the order of $\log_2 n$ bits. If a set of 
integers are to 
be encoded, confusion arises if we pack
all these binary digits together because the receiver does not know where the 
representation of the first digit ends and the second
digit starts. To resolve this, one can also encode the length of the binary 
representation of $n$ in binary representation, whose
length can also be estimated to be $\log_2 \log_2 n$. The same problem arises 
again since the receiver does not know the length of this
preamble representation. This problem can be solved by giving the length of the 
preamble as $\log_2 \log_2 \log_2 n$ as another preamble.
This procedure can be iterated until $\log_2 \ldots \log_2 n$ is as close to zero 
(and positive) as desired. The integer number can be encoded
with a scheme that includes these preambles, so that the total length of the code 
is given by:
\begin{equation}
\label{eq:length_integer}
L(n) = \log^* n = \log_2 c + \log_2 n + \log_2 \log_2 n + \ldots,
\end{equation}
where the constant $c \approx 2.865$ is included for consistency 
\citep[see, e.g.,][]{rissanen89}. The symbol $\log^* n$ is chosen to represent
the previous sum. For large numbers, the dominant factor is $\log_2 n$, so that 
we can approximately assume that the length of the
message for encoding an integer number is of the order of $\log_2 n$ for
sufficiently large $n$ \citep{rissanen83}. For 
encoding integers with sign, we can add a single bit for
setting the sign, so that the length of the message is $\log^* n+1$. The dominant 
factor is again $\log_2 n$.

Encoding real numbers is more complicated than integer numbers because we 
would need an infinite number of bits to encode the real
number to infinite precision. To solve this in practice, we have to encode the real 
number $x$ assuming a 
precision $\delta$, so that the encoded number $x_\delta$ fulfills $|x-x_\delta| < 
\delta$. Once the precision is fixed, we can
encode the integer part $\lfloor x_\delta \rfloor$ and the fractional part of $x_\delta$ 
separately. It can be shown \citep{rissanen78} that the
length of the message is $L(x_\delta)=\log^* \lfloor x_\delta \rfloor + 
\log^*(1/\delta)$. As stated before, if $x$ is large and taking
into account that $x \approx x_\delta$, we can estimate the length of the message 
for encoding a real number $x$ with
\begin{equation}
\label{eq:length_real}
L(x) \approx \log_2 x - \log_2 \delta.
\end{equation}
Note that the length tends to zero when the number to encode approaches the 
precision. In the limit situation that $x$ is smaller than the precision, the
length takes a negative value that does not have meaning.

With the previous considerations in mind, the model selection problem can be 
established. Let $f_i$ be the model that generates the data or the one that is more 
plausible and that it belongs to 
a class of $m$ models $M$. Consider that each
model is parameterized by $k_i$ parameters $\theta_{ij}$, with $1 \leq i \leq m$ 
and $1 \leq j \leq k_i$. Given a set of observations,
our aim is to choose the most plausible model $f_i$ from the set $M$ and to 
estimate the parameters $\{\theta_{ij}, 1 \leq j \leq k_i\}$. The sender splits
up the message
in two parts. The first one contains the model itself and the second one 
contains the departures between the model and the data.
We transmit the information about the model by first encoding the number of 
parameters that characterizes it and then transmitting
the parameters themselves with a given precision $\delta_j$ associated with each 
one:
\begin{equation}
\label{eq:length_n_params}
L(M) = \log^* k_i + \sum_{j=1}^{k_i} \Big[ \log^* k_j + \log^*(1/\delta_j) \Big].
\end{equation}
Since the $\log^* n$ function behaves as $\log_2 n$ for sufficiently large $n$ 
\citep{rissanen83}, we can transform the previous formula to:
\begin{equation}
\label{eq:length_n_params2}
L(M) \approx \log_2 k_i + \sum_{j=1}^{k_i} \Big[ \log_2 k_j - \log_2 \delta_j \Big].
\end{equation}
The previous equation gives the length of the model for encoding the 
parameters with arbitrary precision $\delta_j$. However, it makes no
sense to increase the precision of the parameters unnecessarily because there 
may 
be no information in the data for such a task. It has been
demonstrated by \cite{rissanen89} that if the optimum parameters are computed 
from a large set of $n$ observed data points, the
precision of the parameters can be effectively encoded with only $\onehalf \log_2 
n$ bits. The reason for this is that, if the
parameters are being estimated from the data, it makes no sense to encode them 
with a precision larger than the standard error of the
estimation. For a typical estimation of the $\theta_{ij}$ parameter, the standard 
error decreases with the number of data points
as $1/\sqrt{n}$. Therefore, the code length for encoding a real number with 
precision $1/\sqrt{n}$ is $-\onehalf \log_2 n$. If all the
points are used for estimating all the parameters, we can rewrite Eq. 
(\ref{eq:length_n_params2}) as:
\begin{equation}
\label{eq:length_n_params3}
L(M) \approx \log_2 k_i + \sum_{j=1}^{k_i} \log_2 k_j + \frac{k_i}{2} \log_2 n.
\end{equation}

Additionally, we have to encode the observed data given the model to obtain the 
length $L(D|M)$. In the simple case in which we do not
have any information about
the probability distribution of the data given the model, we can apply the same
encoding scheme we have applied for describing the model.
Assume that we have $n$ observed data points $y_k$ and that the model $f$ 
produces a fit to the observations such that the residual can
be written as:
\begin{equation}
\label{eq:residual}
r_j = | y_j - f(x_j)|.
\end{equation}
The encoding is performed by saving the number of data points and the value of 
each data point by using Eq. (\ref{eq:length_real}). The length $L(D|M)$ is:
\begin{equation}
\label{eq:length_data}
L(D|M) = \log^* n + \sum_{j=1}^{n} \Big[ \log^* r_j + \log^* (1/\delta_j) \Big],
\end{equation}
which can be simplified if the number of data points $n$ is large enough to give:
\begin{equation}
\label{eq:length_data2}
L(D|M) \approx \log_2 n + \sum_{j=1}^{n} \Big[ \log_2 r_j - \log_2 \delta_j \Big].
\end{equation}
Putting together Eqs. (\ref{eq:length_n_params}) and (\ref{eq:length_data}), the 
total 2-part length of the message is:
\begin{eqnarray}
\label{eq:length_total}
L &=& \log^* k_i + \sum_{j=1}^{k_i} \Big[ \log^* k_j + \log^*(1/\delta_j) \Big] 
\nonumber \\
&+& \log^* n + \sum_{j=1}^{n} \Big[ \log^* r_j + \log^* (1/\delta_j) \Big].
\end{eqnarray}
If the number of data points $n$ is large enough, we can use 
Eqs. (\ref{eq:length_n_params2}) and (\ref{eq:length_data2}) to obtain:
\begin{eqnarray}
\label{eq:length_total_approximate}
L &\approx& \log_2 k_i + \sum_{j=1}^{k_i} \Big[ \log_2 k_j - \log_2 \delta_j \Big] 
\nonumber \\
&+& \log_2 n + \sum_{j=1}^{n} \Big[ \log_2 r_j - \log_2 \delta_j \Big].
\end{eqnarray}

\subsection{Computer-oriented Code Lengths}
\label{sec:computer_oriented}
The previous equations are considered for an optimum encoding scheme. 
However, a simpler and more computer-oriented estimation can
also be obtained based on the previous results with good results \citep{gao00}. 
We assume that 
integer numbers are saved in a computer using
$l_\mathrm{I}$ bits, while real numbers are saved in the memory of the computer 
using $l_\mathrm{R}$ bits. Usually, integer
numbers are saved using $l_\mathrm{I}=16$ bits, while real numbers can be 
stored using either $l_\mathrm{R}=32$ or
$l_\mathrm{R}=64$ bits depending on the desired precision. Therefore, we can 
estimate the length of the message that describes the
model (the number of bits required for its storage in the memory of the computer) 
by:
\begin{equation}
\label{eq:length_bits_model}
L(M) \approx k_i l_\mathrm{I},
\end{equation}
where the number of bits required for transmitting the number of parameters given 
by $log^* k_i$ can be usually neglected with respect
to the length of the message used for transmitting the parameters themselves.
Concerning the storage of the data, we consider that the model output is correct 
when the residual is smaller than a given precision
threshold $\delta$. If the model output is incorrect, we store the difference as a 
real number with $l_\mathrm{R}$ bits, so that
the length is given by:
\begin{equation}
\label{eq:length_bits_data}
L(D|M) \approx \sum_{r_j > \delta} l_\mathrm{R}.
\end{equation}
The length of the parameters associated with the model increase linearly with the 
number of parameters, while the length of the data set
decreases, with a slope that depends on how well the model fits the data.

\subsection{Known distribution}
The previous encoding schemes have been obtained using universal priors, thus 
assuming no knowledge at all about the distribution
of the values of the data and/or model. However, if some information about the 
probability distribution is known, it can be
incorporated in the description of the encoding length through Eq. 
(\ref{eq:shannon}). A typical case is when the observed data
is contaminated with noise described by a Gaussian distribution with zero 
mean and a given standard deviation $\sigma$ (that may even be unknown).
For the set of $n$ observations $y_j$, the probability distribution of the residuals is 
given by:
\begin{equation}
P(\mathbf{r}) = \prod_{j=1}^n \left( 2 \pi \sigma_j^2 \right)^{-1/2} \exp 
\left[ - \frac{r_j^2}{2 \sigma_j^2} \right],
\end{equation}
where $\mathbf{r}$ is the vector of $n$ residual $(r_1,r_2,\ldots,r_n)$ and 
$\sigma_j^2$ is the variance for each data point. 
Assuming the same variance $\sigma^2$ for all the data points and using 
Shannon's theorem, we obtain:
\begin{equation}
L = \frac{n}{2} \log \left( 2  \pi \sigma^2 \right) + \frac{RSS}{ 2 \sigma^2},
\end{equation}
where $RSS= \sum_{j=1}^n r_j^2$ is the residual sum of squares.
The previous equation constitute the estimation of the length for communicating
the data set when we have an estimation for the value of $\sigma$. If the
variance is not known \citep{rissanen89}, we can use the maximum likelihood 
estimation $\sigma^2 \approx RSS / n$ and obtain:
\begin{equation}
\label{eq:length_gaussian}
L \approx \frac{n}{2} + \frac{n}{2} \log \frac{2  \pi}{n} + \frac{n}{2} \log 
RSS,
\end{equation}

\subsection{Example}
\begin{figure*}[!ht]
\plottwo{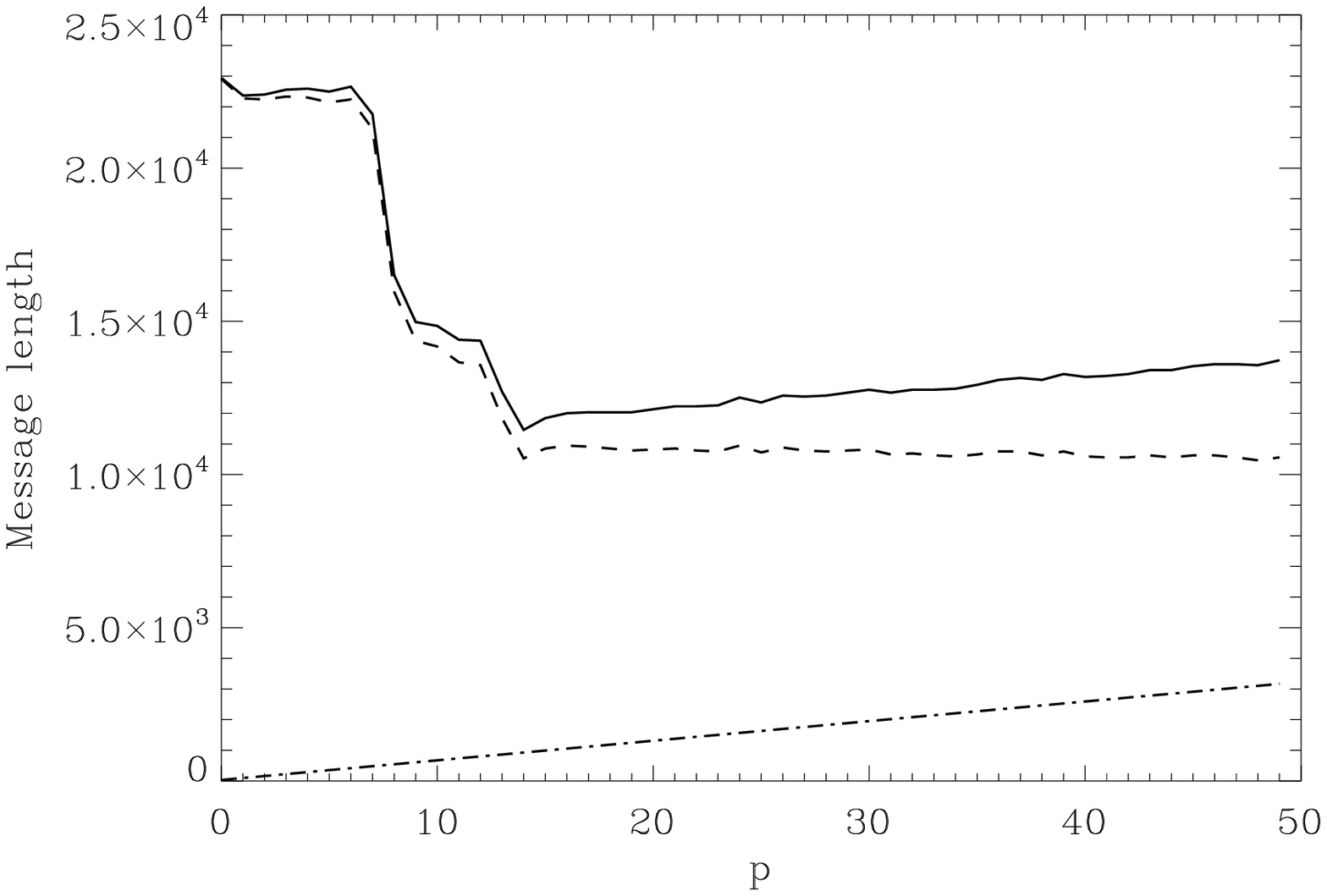}{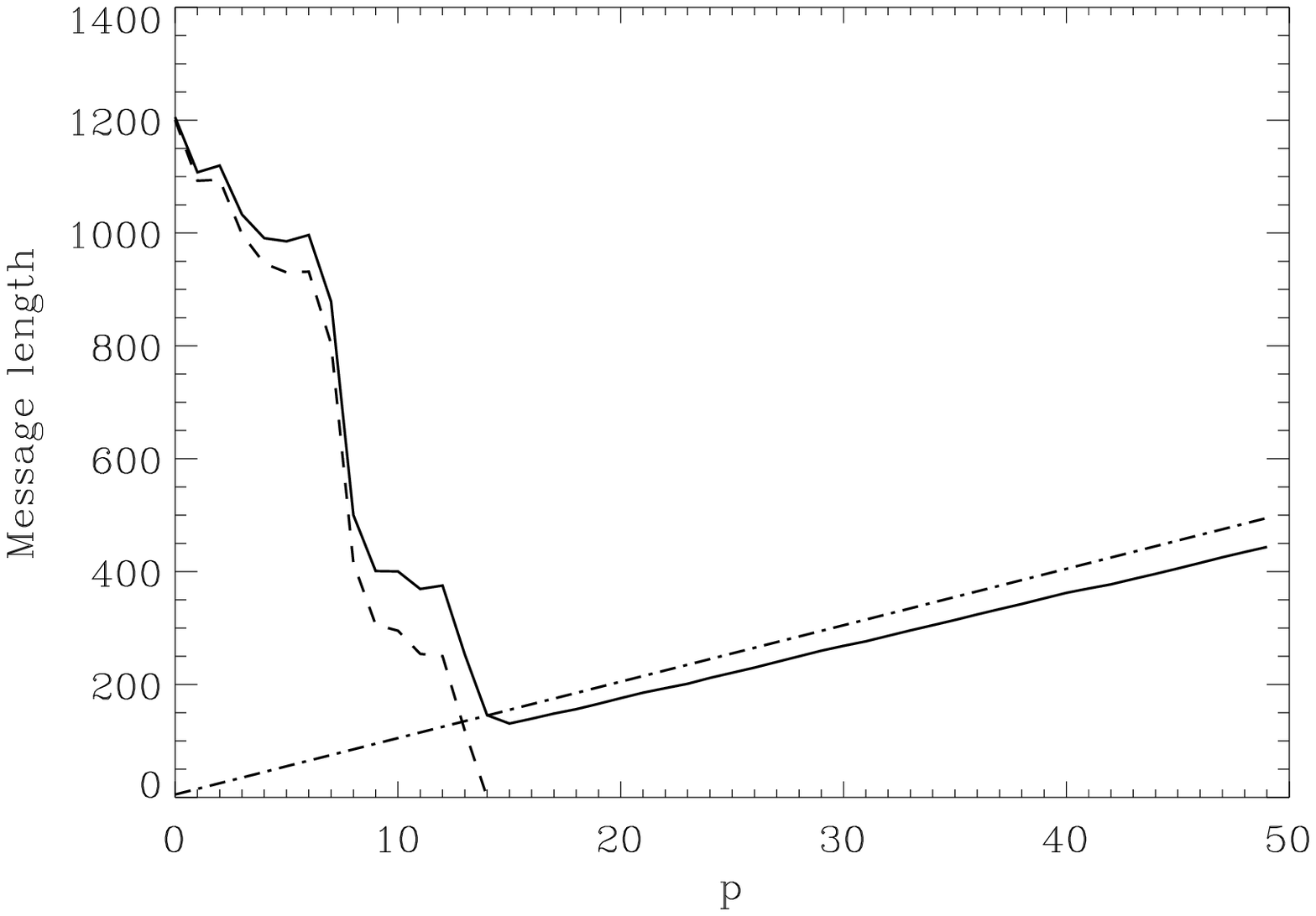}
\caption{Example showing the application of the MDL criterion for model
selection. The aim is to fit a linear combination of sinusoidal signals with 
maximum frequency 14 using a Fourier series. The left panel shows the message 
length obtained using
the computer-oriented code lengths. The right panel shows the results
obtained using a gaussian distribution for the probability density of the
residual. Note that in both cases we find that MDL criterion gives the correct
value of the maximum frequency.\label{fig:fourier}}
\end{figure*}

For demonstrating the previous machinery in a simple practical problem,
let us assume that we have a noisy linear combination of sinusoidal signal:
\begin{eqnarray}
\label{eq:signal_original}
f(x) &=& 2 \sin(\pi x) + \sin(3\pi x) - \sin(4\pi x) \nonumber \\
&-& 2\cos(8\pi x) + \cos(14\pi x) + \epsilon.
\end{eqnarray}
The variable $x$ is always inside the interval $[0,1]$ and $\epsilon$ is a noise 
term with zero mean and standard 
deviation $\sigma$. Let us consider that we have sampled
the $x$ axis in 512 points and that $\sigma = (\max(f)-\min(f))/8$. The 
frequencies of the signals
can be obtained by performing the Fast Fourier Transform of the signal. However, 
for applying the MDL principle, we consider models of the type
\begin{equation}
\label{eq:signal_fourier}
\hat f(x) = a_0 + \sum_{j=1}^p \left\{ a_j \cos(j \pi x) + b_j \sin(j \pi x) \right\}.
\end{equation}
The aim is to obtain the most plausible value of $p$ that produces the smallest 
encoding length of the model and data
given the model. Firstly, we apply the code length described in 
\S\ref{sec:computer_oriented} using a threshold equal to the standard
deviation of the data. The results are shown in the left panel of Fig. \ref{fig:fourier}. 
Note that the number of bits necessary to
encode the model increases linearly. The number of bits to describe the 
residual between the model and the observed data
decreases rapidly when $p \lesssim 10$. The sum of both terms has a minimum 
around $p \sim 14-15$, which coincides with the maximum
value of $p$ in the original signal.

The right panel shows the results obtained when we take into account that the 
residuals between the model and the data are well characterized by
a Gaussian distribution and that the number of points is sufficiently large. In 
this case, the length of the model is given by
Eq. (\ref{eq:length_n_params3}), while the length of the data is given by 
Eq. (\ref{eq:length_gaussian}). The 
minimum of the total length gives the most
plausible value of $p \sim 14-15$, compatible with the original data and with the 
value given by the previous estimation of the total
encoding length.

\section{Applications}
In this section we apply the MDL principle to two selected problems in
the field of solar spectropolarimetry in order to show the potential of 
this technique for model selection. Our aim is to apply the MDL criterion
systematically to other similar problems in the future.

\subsection{PCA de-noising}
As a first application, we consider the case of Principal Component Analysis
(PCA) de-noising of spectropolarimetric 
observations. The dataset consists on full-Stokes 
observations of the two \ion{Fe}{1} at 15648 \AA\ and 15652 \AA\ in an
extremely quiet region of the solar surface. This dataset has been used by
\cite{marian06} to investigate the magnetic properties of the quiet Sun. The
signal-to-noise ratio of the observations have been improved by using the
PCA de-noising technique.

PCA is a statistical technique that, given a data set, produces a set of orthogonal 
vectors and eigenvalues that can be used for decomposing the original data.
These eigenvectors point in the directions of maximum covariance. The 
eigenvector associated with the largest eigenvalue points along the direction with 
the largest
covariance in the data and so on. PCA provides an ``optimal''
basis set for decomposing (and reconstructing) the data set. Alternatively,
it has been used for compressing data (by saving only the eigenvalues and 
eigenvectors) or for efficiently inverting Stokes profiles \citep{rees_PCA00}.
For computing the 
PCA decomposition, 
the covariance matrix of the observations has to be diagonalized. Once the 
eigenvectors $\mathbf{e}^i(S)$ (vectors whose dimension $N_\lambda$ equal the 
number of
wavelenghts in the dataset) and eigenvalues $\lambda_i(S)$ of the covariance 
matrix associated with the Stokes parameter $S$ are obtained, any of the Stokes 
profiles can be decomposed as:
\begin{equation}
\label{eq:PCA_summation}
S(\lambda_j) = \sum_{i=0}^N \lambda_i(S) e^i_j(S),
\end{equation}
where the subindex $j$ refers to the wavelength position. The previous equation
can be alternatively seen as a technique for reconstructing the original signal
from the eigenvalues and eigenvectors obtained after the PCA decomposition.
The precision of the 
reconstruction can be modified by changing the value of the number of 
eigenvectors
included ($N$). The PCA reconstruction assures that the error of the 
reconstruction decreases when more eigenvectors are included. The 
eigenvectors with the largest eigenvalues represent the features that are more
statistically representative of the dataset, while the noise and particular features
are accounted for by the eigenvectors with the smallest eigenvalues. 
As a consequence, it is possible to get rid of the majority of the noise in the
observation by stopping the summation of Eq. (\ref{eq:PCA_summation})
in a suitable $N' \leq N$. Nevertheless, it is not an easy task to find a criterion for 
selecting this $N'$ because instrumental effects plus data reduction can introduce
spurious signals that have some kind of correlation. As a consequence, they
contribute to the eigenvectors that carry the relevant polarimetric information.

\begin{figure*}
\plottwo{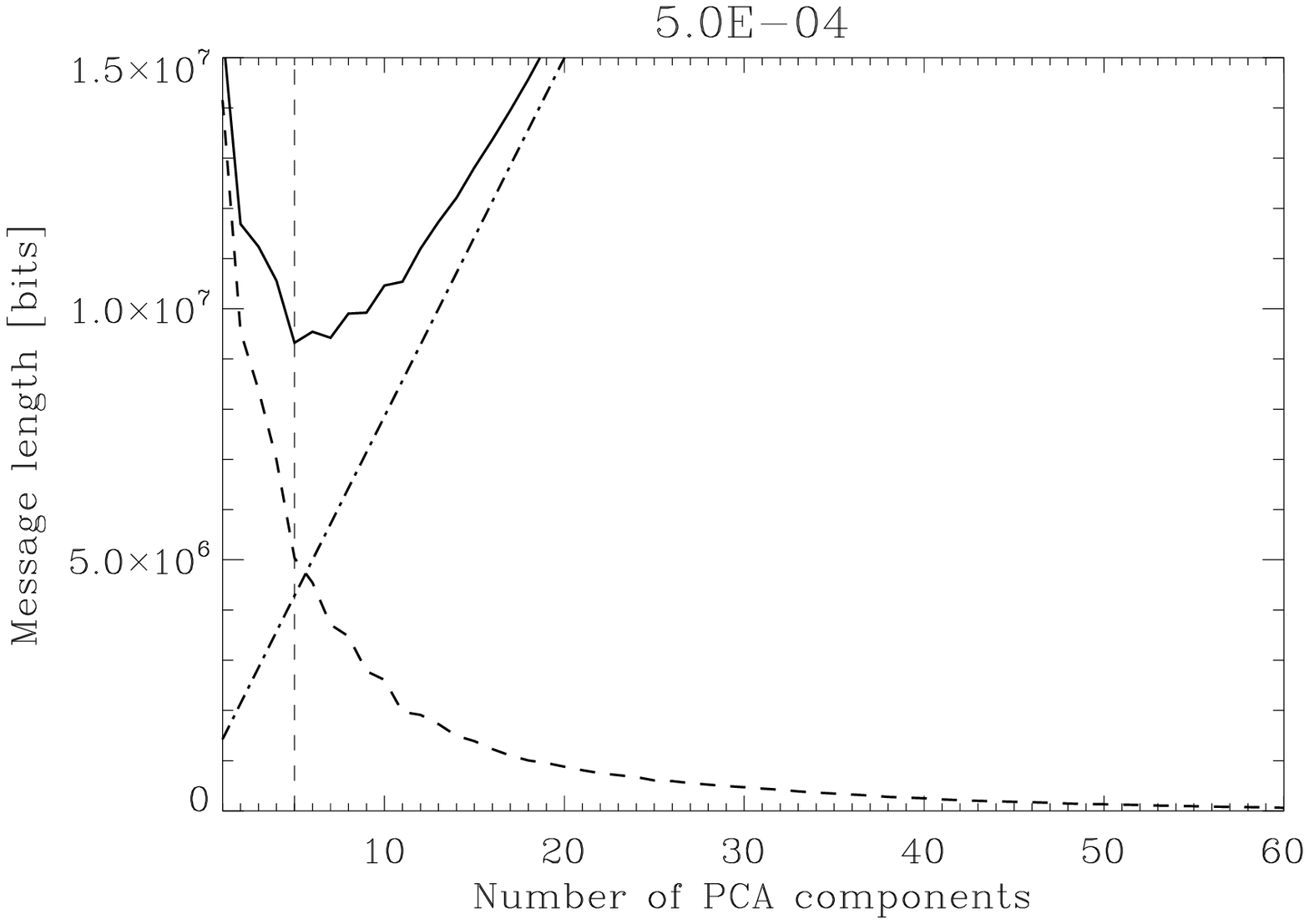}{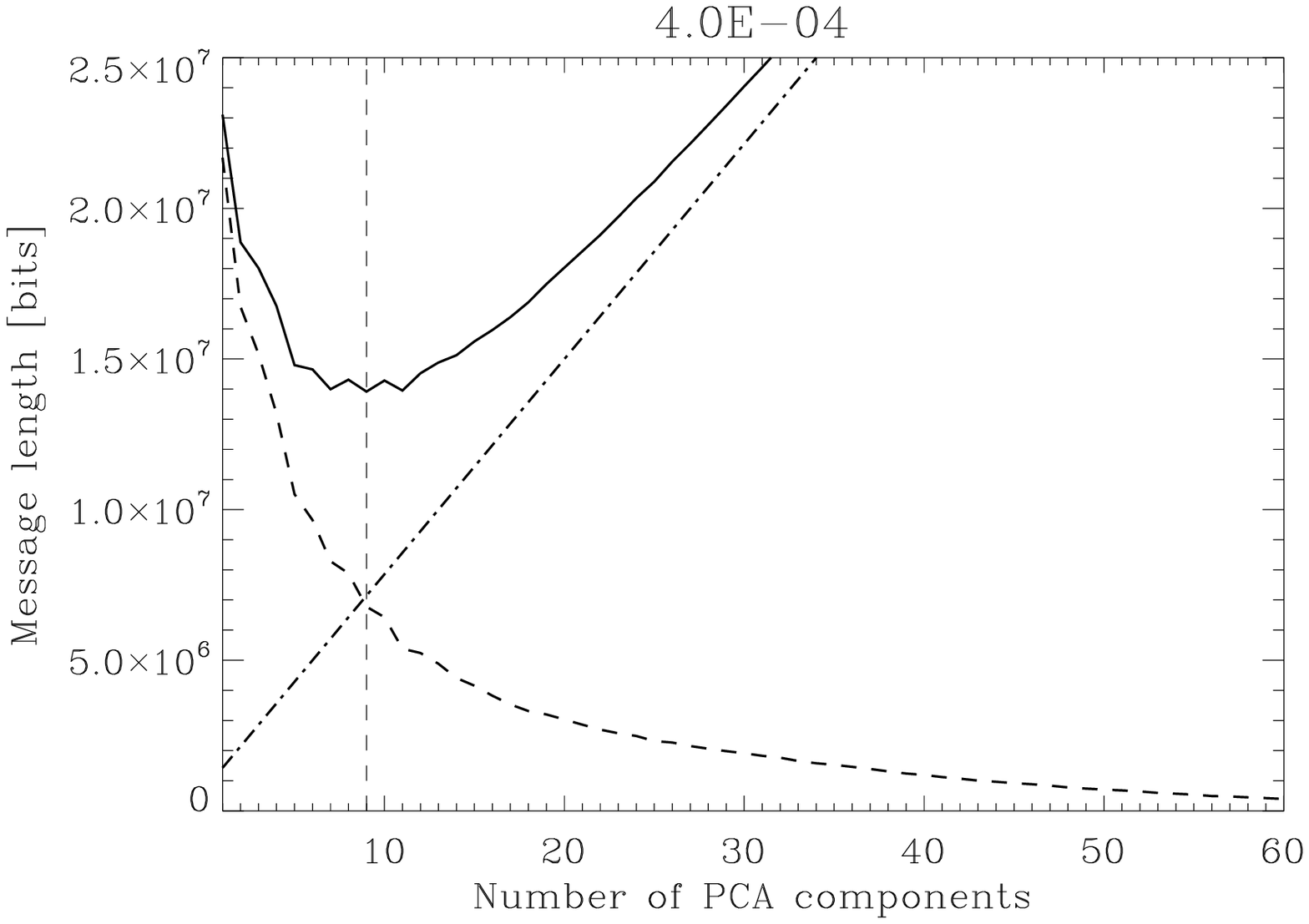}
\plottwo{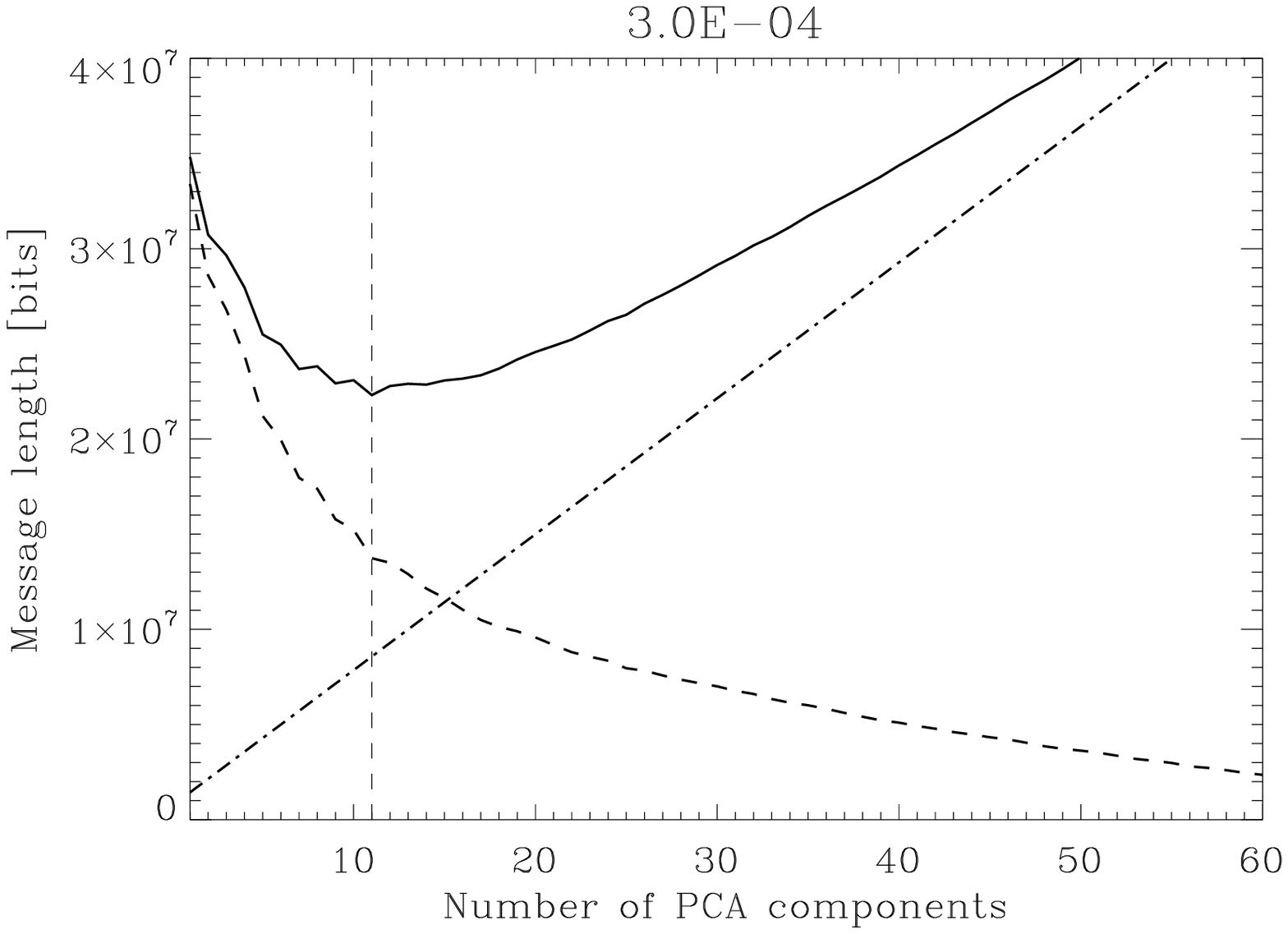}{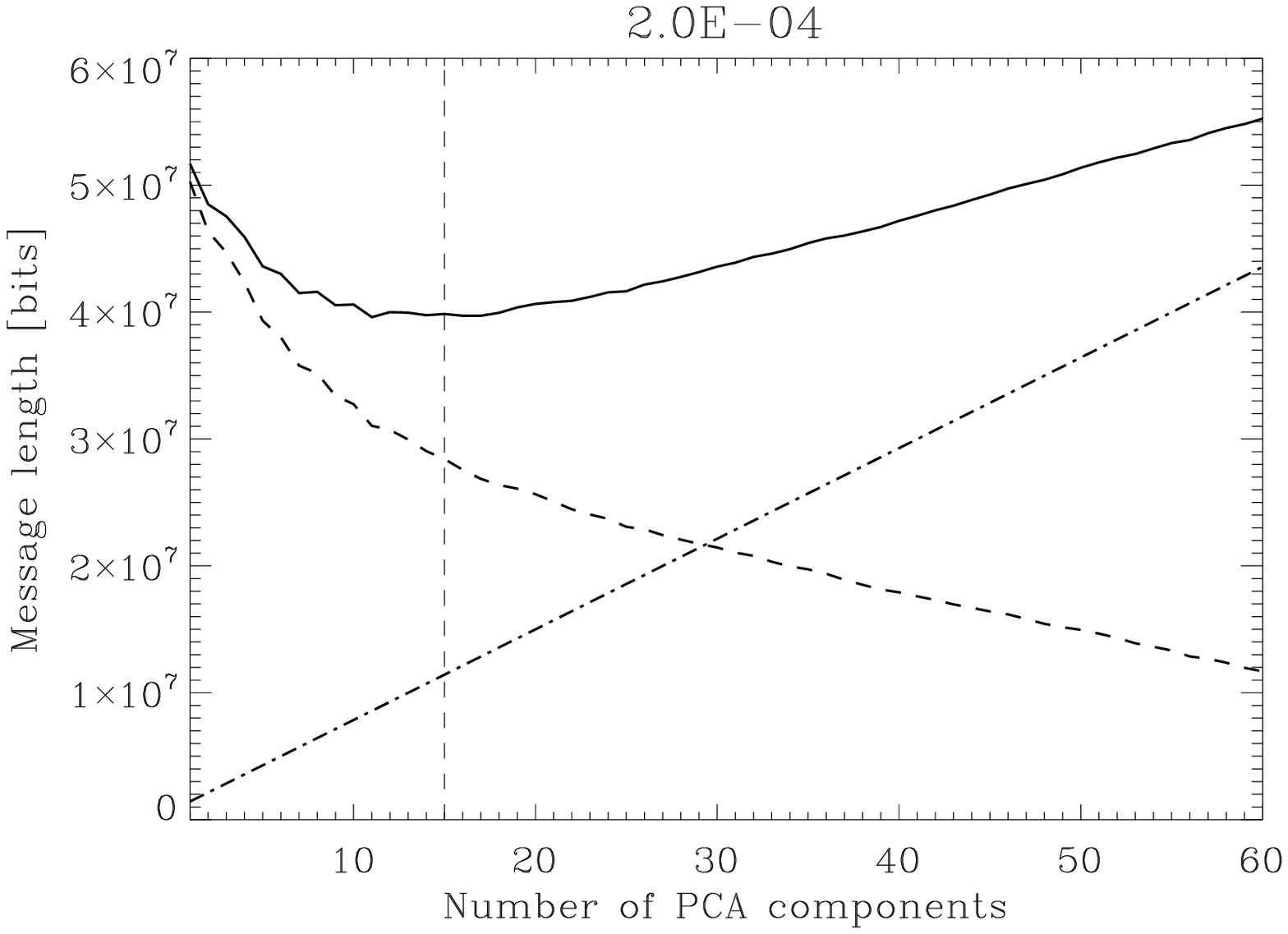}
\caption{Application of the MDL criterion to the de-noising of spectropolarimetric
signals of quiet Sun observations. The length of the model (dot-dashed line), the 
data set given the model (dashed line) and the total length (solid line) are plotted
versus the number of PCA components included in the data reconstruction
for different values of the threshold that sets the precision of the reconstruction
(shown in the title of each plot). The vertical dashed line indicates the approximate
minimum of the total length curve. Note that the number of PCA components 
obtained with this MDL criterion increases as the threshold decreases.
\label{fig:PCA}}
\end{figure*}

We propose to use the MDL criterion to select this optimal $N'$. The experiment
is carried out with the Stokes V profiles of the \ion{Fe}{1} lines at 15648 \AA\ and 
15652 \AA\
observed in a very quiet internetwork region of the Sun 
described by \cite{marian06}. The original data presents a signal-to-noise 
ratio (SNR)
of $\sim$5 for the 15648 \AA\ line and $\sim$2 for the 15652 \AA\ line.
The summation
of Eq. (\ref{eq:PCA_summation}) is calculated for increasing values of $N$ and
the MDL length is calculated using the technique described in 
\S\ref{sec:computer_oriented}. The length of the model $L(M)$ is obtained by 
calculating the number of bits to represent the eigenvectors $\mathbf{e}^i$ with
$i=0,\ldots,N$ plus the coefficients $\lambda_i$ for expanding all the profiles
in the field of view. The length of the data set given the model $L(D|M)$ is obtained 
by calculating the number of bits needed for encoding the reconstructed profiles 
that differ from the original profiles by more than a given threshold. 
Figure \ref{fig:PCA} shows these lengths versus $N$ for different values of this
threshold. The
length of the model is plotted in dot-dashed lines, the length of the data in
dashed line and the total length in solid line. We have also marked the value of
$N$ at which we obtain the minimum of the total length. This is the MDL 
optimum value $N'$ for the number of eigenvectors. Note that this minimum
increases
when the allowed threshold decreases, a consequence of putting more restrictions 
to the model. Of course, this threshold should be chosen consistent with the 
expected noise in the observations.

\subsection{LTE inversion}
The diagnostic of magnetic fields via the interpretation of spectropolarimetric
observations is often based on the assumption of a model. Sometimes, the 
observations themselves do not carry enough information for discriminating 
among several models. A similar problem arises when a model can have 
an arbitrarily large number of parameters and it is not an easy task to select
an optimum value of such parameters. A commonly used technique to minimize
over-fitting is to use models with as few parameters as possible. Nevertheless,
the data may contain enough information for constraining more parameters
and we may be using overly simple models to interpret the observations.

We propose to use the MDL criterion to discriminate among different models
that can be used to describe a set of observations. Our aim is to introduce
in the community an easy technique for confronting different models when
applied to the same data set. In the framework of the MDL criterion, the 
researcher is able to objectively discriminate one of the models among 
the others, making sure that the selected model explains the observations without 
over-fitting them.

\begin{figure*}[t]
\plottwo{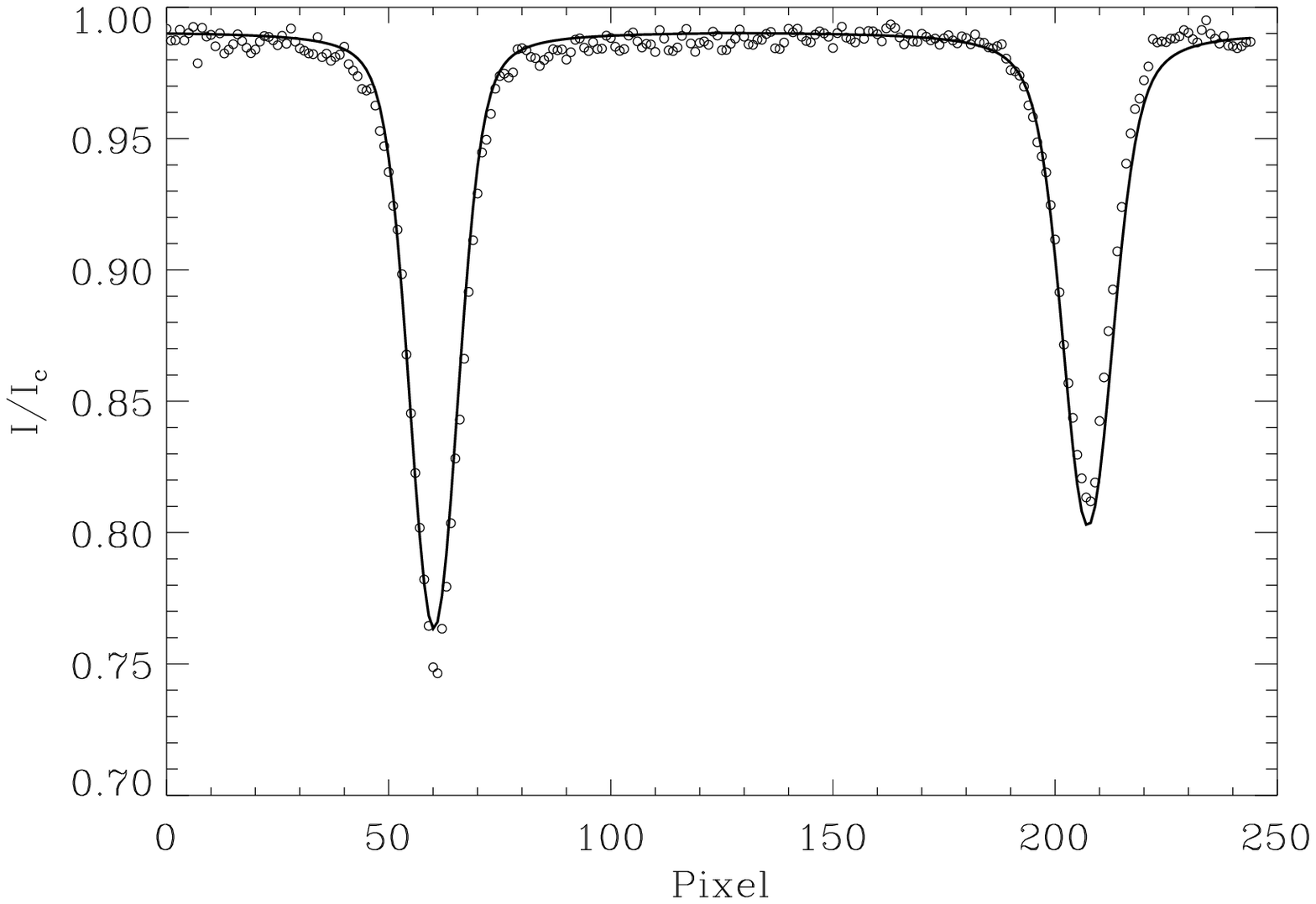}{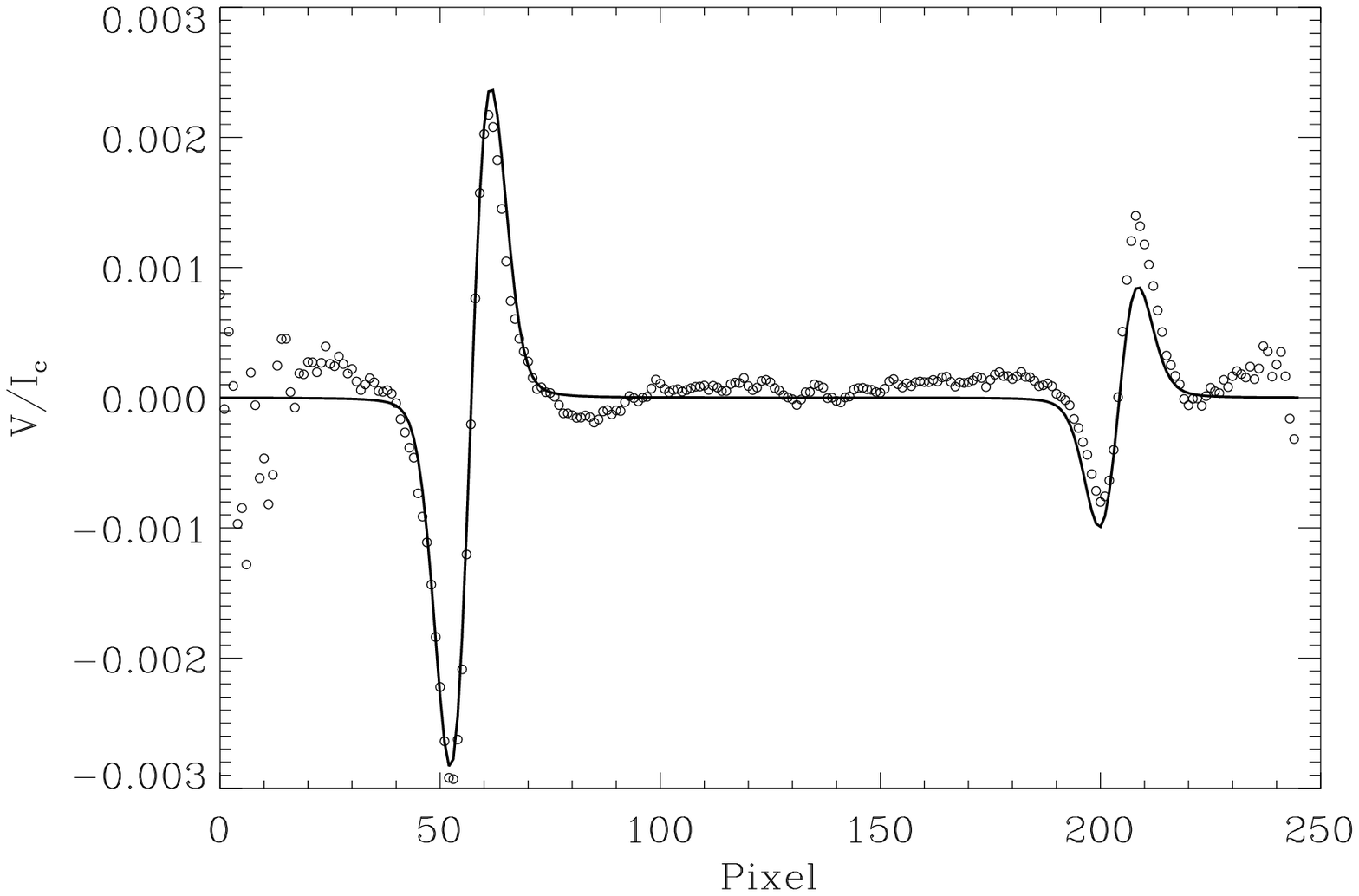}
\caption{One of the observed Stokes I (left panel) and Stokes V (right panel) 
profiles is shown in circles. The fits obtained with the two-component
model SIR inversion are shown in solid lines. Note that this model is not
powerful enough for fitting this strongly asymmetric profiles. Both fits have
been obtained using 2 nodes in the temperature. The fit of the Stokes V 
profiles has been obtained using 4 nodes in the magnetic field strength
depth profile.\label{fig:observations}}
\end{figure*}

We demonstrate our approach by using the LTE inversion code SIR 
\citep[Stokes Inversion based on Response functions;][]{sir92}. An 
example of the
Stokes profiles of the \ion{Fe}{1} lines at 15648 \AA\ and 15652 \AA\ observed
in a very quiet internetwork region are shown in Fig. \ref{fig:observations} 
\citep{marian06}. The inversion is carried out 
with a two-component
model (a magnetic one occupying a fraction of the resolution element and
a non-magnetic one filling up the rest of the space).
The observations clearly show strongly distorted Stokes profiles 
that cannot be correctly
reproduced with this simple two-component model. Although very simple,
this test demonstrates the capabilities of the MDL criterion to pick up a model
when none of the models of the proposed set is able to correctly fit the 
observations. SIR represents the variation with depth of the thermodynamical
and magnetic properties with the aid of nodes that are equidistant
in the $\log \tau$ axis. It is of interest to note that when a new node is 
included for representing the depth variation of a physical quantity, the 
previous nodes are shifted so that the final distribution is again equidistant.
Splines are used to 
interpolate these quantities between the nodes.
The number of nodes of the temperature (or the magnetic field
strength) is increased and the MDL criterion is used for
selecting the optimal values. We follow the prescriptions 
proposed in \S\ref{sec:computer_oriented}. The length of the model is 
calculated as the number of bits to encode the number of parameters
plus their values. The length of the data given the model is chosen to be
equal to the number of bits neccesary to encode the points in the profile
for which the relative difference between the model and the data is above a
certain threshold. This threshold represents the tolerance we allow
in our model for considering that it fits our observations.

\begin{figure*}[t]
\plottwo{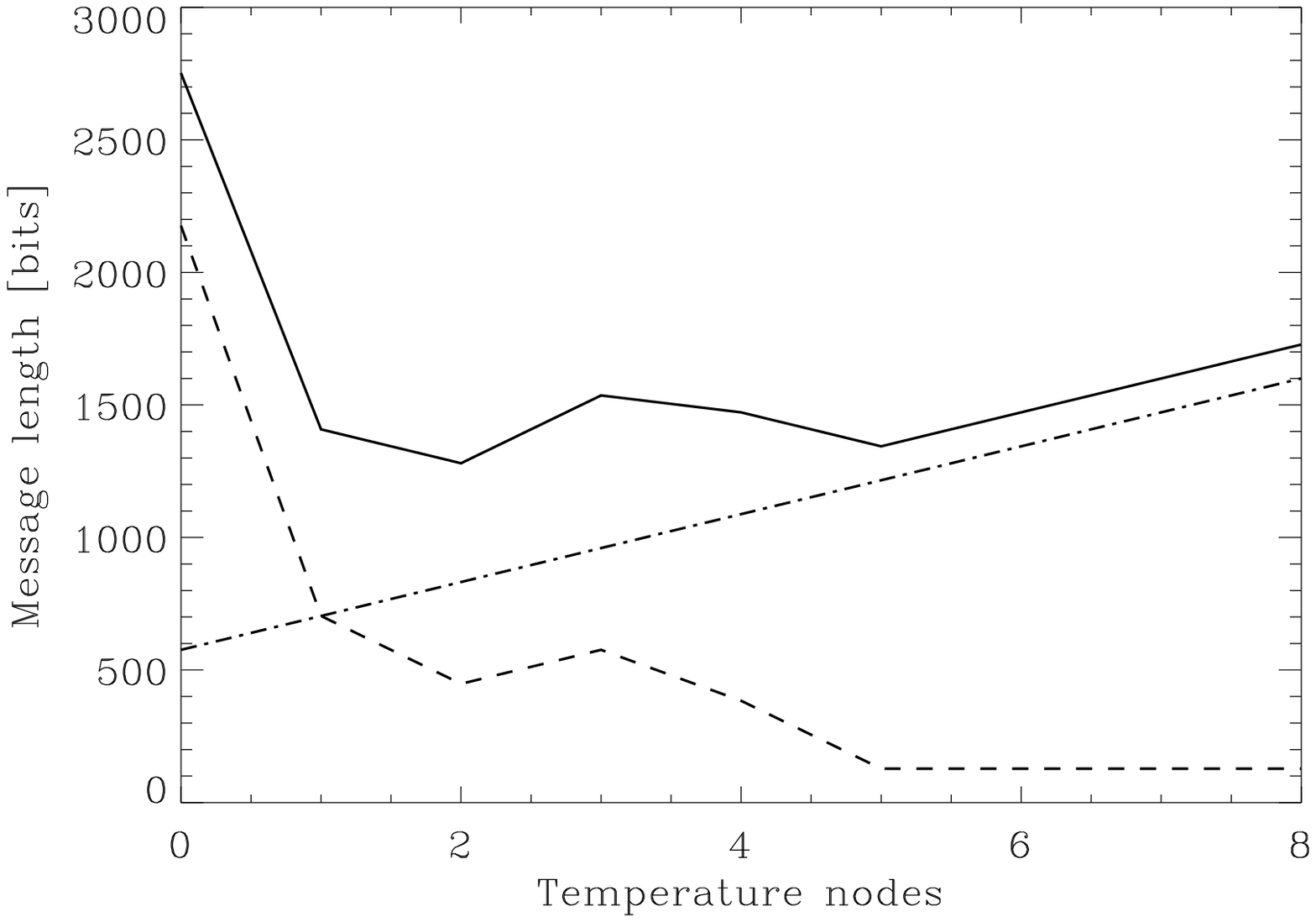}{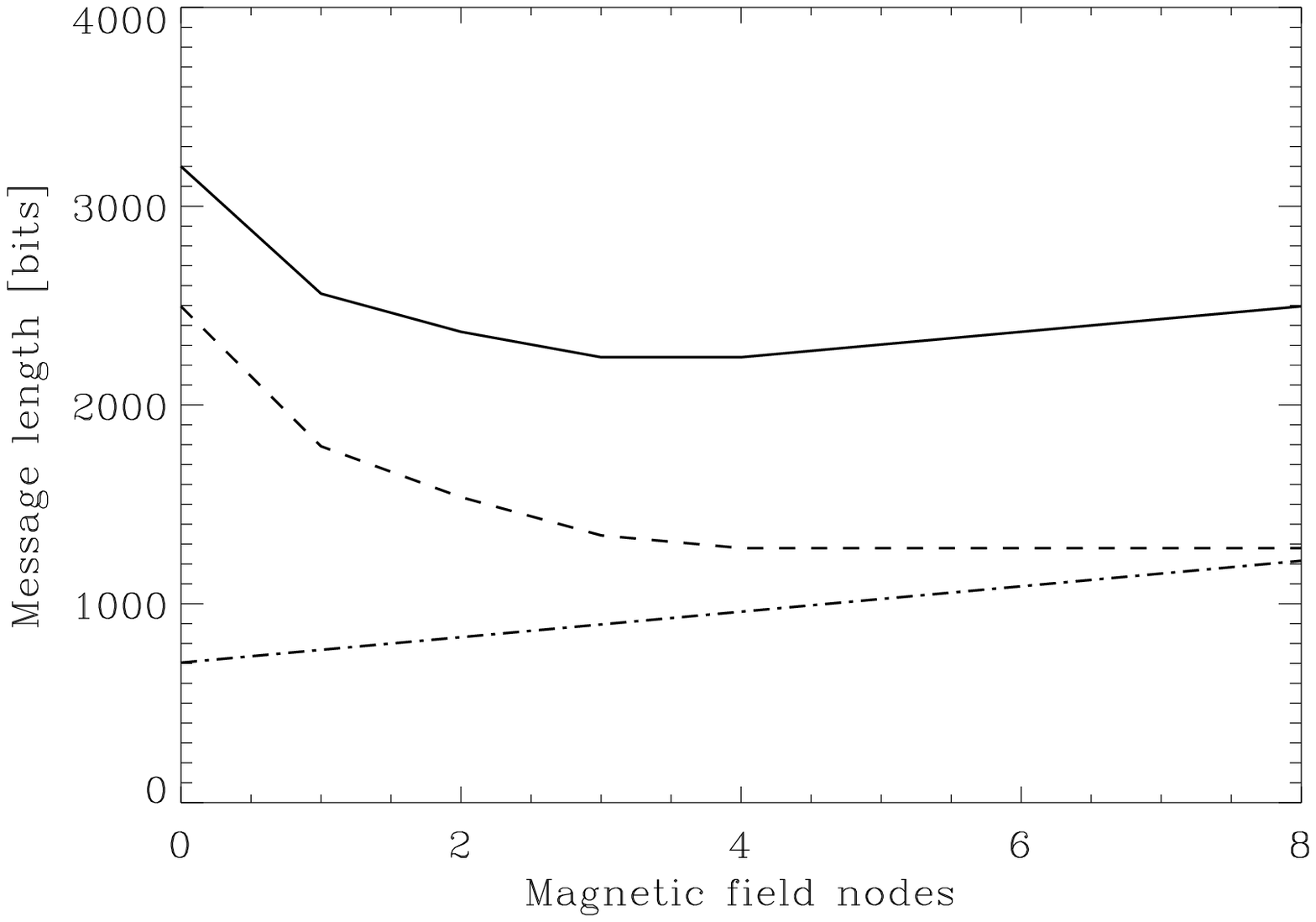}
\caption{Application of the MDL criterion to the selection of the optimum number 
of nodes in an LTE inversion with the SIR code. The left panel shows the 
message length when only the number of nodes of the temperature depth profile 
is changed. The right panel shows the message length when the number of 
nodes of the magnetic field strength profile is changed.\label{fig:LTE}}
\end{figure*}

The first experiment consists in selecting the optimal number of nodes in the 
temperature when
the number of nodes in the rest of variables are kept constant. We only 
compare
the Stokes I observed and synthetic profiles in this experiment because it is the 
only Stokes 
parameter that is almost unsensitive to the magnetic properties of the
atmosphere. This is not generally the case since Stokes I can be also sensitive to 
the
magnetic field in the strong field regime. In analysing internetwork quiet Sun 
Stokes profiles, the 
filling factor of the magnetic component needs to be very small. In such a
situation, the emergent Stokes I profile fundamentally depends on the properties of
the non-magnetic component, while the emergent Stokes V profile depends
only on the properties of the magnetic component.
We consider that a point in the profile reproduces the observations
if the relative error between the observed and the synthetic profiles is below 2\%.
The results for the message length are shown in the left panel of Fig. \ref{fig:LTE},
while the fit is shown in the left panel of Fig. \ref{fig:observations}. The MDL 
criterion demonstrates that $\sim$2 nodes are required
in the temperature depth profile. Fewer nodes give a fit to the Stokes profiles that 
is too bad. More nodes give a model that takes more bits to communicate than
the uncorrectly fitted points themselves.

The second experiment is similar, but in this case we select the number of
nodes of the magnetic field strength. The threshold for the relative error in
the Stokes I profiles is 2\%, while we increase it to 10\% for the Stokes V
profiles. This large relative error for the Stokes V profile is motivated by the 
poor fit we obtain to the observed Stokes profiles with the simple 
two-component model. The message length is shown
in the right panel of Fig. \ref{fig:LTE}, while the fit is shown in the right panel
of Fig. \ref{fig:observations}. The optimum number of nodes for the
magnetic field suggested by the MDL criterion is $\sim$3.

In this section we have presented a very simple problem concerning the 
selection of the optimal number of parameters in an LTE inversion. 
However, we consider that this approach will be of great interest for
solving such a difficult problem in a simple way and we suggest calculating
the MDL criterion for all the fits performed to an observed profile. 

\section{Conclusions}
We have presented the Minimum Description Length Principle developed by 
\cite{rissanen78} to discriminate between a set of available models that can
approximate a given data set. We have briefly presented its relation with 
the Bayesian approach of model selection through the application of the
Shannon's theorem. In our opinion, the MDL principle presents a user
friendly procedure for model selection. We have presented simple ways of
estimating the message length that can be applied for communicating
integer and real numbers. A more computer-oriented and easy to implement
procedure has been also shown.

For the sake of clarity, we have applied the MDL principle to simplified problems.
The selection of the optimal number of PCA components when de-noising 
observed Stokes profiles and the selection of the optimal number of nodes
in the temperature and magnetic field depth profile obtained from LTE inversion
of Stokes profiles. The results show the potential of this technique for model
selection, with the advantage of being very simple to calculate.

As the main conclusion of this paper, we propose using the MDL principle
as a way to quantitatively select among different competing models.
We propose to include the description length as one of the final outputs of any 
inversion code. This makes it very easy to select the optimal model from
a proposed set of models based on the framework of the MDL principle. 
It is of interest to stress that this principle can be used to select among
models that are based on different scenarios. Usually, most complex
scenarios translate into more degrees of freedom that may not be
constrained by the observations. Using the MDL principle, the
selected model among all the possibilities might not give the
best fit to the observations but it represents the model that produces
a good fit with a conservative number of parameters.

As a final remark, a possible future way of research may be how 
the MDL principle can be implemented as a 
regularization of the merit function in the existing inversion 
techniques. As a consequence, one would end up with an inversion
code that automatically selects the optimal model.

\acknowledgements
We thank M. J. Mart\'{\i}nez Gonz\'alez for helpful discussions on the
subject of the paper. We also thank Thornsten A. Carroll for his
careful review and useful comments.
This research has been funded by the Spanish Ministerio de Educaci\'on y 
Ciencia through project AYA2004-05792.



\begin{figure}
\plotone{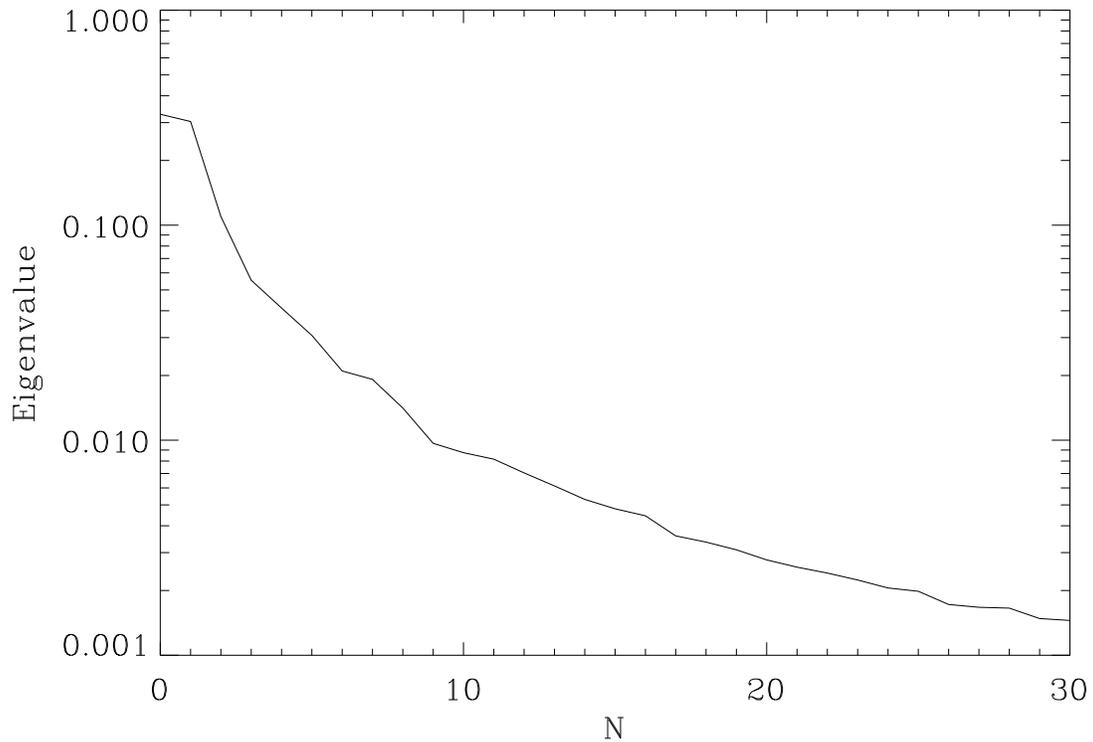}
\caption{Eigenvalues obtained from the decomposition of the observed Stokes V
profiles described by \cite{marian06}. Note the monotonic decay. The eigenvectors associated
with the largest eigenvalues carry most of the signal (features that present strong
correlations for a large set of observed profiles in the field-of-view), while the eigenvectors
associated with the smallest eigenvalues are mainly associated with uncorrelated
noise.
\label{fig:eigenvalues}}
\end{figure}

\end{document}